\documentclass[12pt]{article}

\usepackage{epsfig,multicol,multirow,cite}
\usepackage{amsmath,subfigure,latexsym,amssymb}
\usepackage{hyperref}
\usepackage[figuresright]{rotating}
\usepackage{tikz, graphicx}

\newcommand{\be}{\begin{equation}}
\newcommand{\ee}{\end{equation}}

\newcommand{\bea}{\begin{eqnarray}}
\newcommand{\eea}{\end{eqnarray}} 

\newcommand{\la}{\langle}
\newcommand{\ra}{\rangle}

\newcommand{\Z}{\mathbb{Z}}
\newcommand{\R}{{\kern+.25em\sf{R}\kern-.78em\sf{I} \kern+.78em\kern-.25em}}
\newcommand{\RR}{{\kern+.25em\sf{R}\kern-.6em\sf{I} \kern+.6em\kern-.25em}}
\newcommand{\N}{{\kern+.25em\sf{N}\kern-.78em\sf{I} \kern+.78em\kern-.25em}}
\newcommand{\C}{{\kern+.25em\sf{C}\kern-.50em\sf{I} \kern+.50em\kern-.25em}}

\newcommand{\gtapprox}{\raisebox{-0.5ex}{$\,\stackrel{>}{\scriptstyle\sim}\,$}}
\newcommand{\ltapprox}{\raisebox{-0.5ex}{$\,\stackrel{<}{\scriptstyle\sim}\,$}}

\usepackage{color}
\definecolor{col_red}{rgb}{1.0,0.0,0.0}

\makeatletter
\@addtoreset{equation}{section}
\makeatother

\begin{document}
 
\begin{center}
{\Large\bf An analogue to the pion decay constant}

\vspace*{6mm}

{\Large\bf in the multi-flavor Schwinger model} \\

\vspace*{1cm}

Jaime Fabi\'{a}n Nieto Castellanos$^{\rm a}$,
Ivan Hip$^{\rm b}$ and Wolfgang Bietenholz$^{\rm a}$\\
\ \\
$^{\rm \, a}$ Instituto de Ciencias Nucleares \\
Universidad Nacional Aut\'{o}noma de M\'{e}xico \\
A.P. 70-543, C.P. 04510 Ciudad de M\'{e}xico, Mexico\\
\ \\ \vspace{-3mm}
$^{\rm \, b}$ University of Zagreb Faculty of Geotechnical Engineering \\
Hallerova aleja 7, 42000 Vara\v{z}din, Croatia

\end{center}

\vspace*{6mm}

\noindent
We study the Schwinger model with $N_{\rm f} \geq 2$ degenerate fermion
flavors, by means of lattice simulations. We use dynamical Wilson
fermions for $N_{\rm f} = 2$, and re-weighted quenched configurations
for overlap-hypercube fermions with $N_{\rm f} \leq 6$.
In this framework, we explore an analogue of the QCD pion decay
constant $F_{\pi}$, which is dimensionless in $d=2$, and which
has hardly been considered in the literature.
We determine $F_{\pi}$ by three independent methods, with numerical
and analytical ingredients.
First, we consider the 2-dimensional version of the
Gell-Mann--Oakes--Renner relation, where we insert both theoretical
and numerical values for the quantities involved.
Next we refer to the $\delta$-regime, {\it i.e.}\ a small spatial
volume, where we assume formulae from Chiral Perturbation Theory to
apply even in the absence of Nambu-Goldstone bosons. We further
postulate an effective relation between $N_{\rm f}$ and the number
of relevant, light bosons, which we denote as ``pions''.
Thus $F_{\pi}$ is obtained from the residual ``pion''
mass in the chiral limit, which is a finite-size effect.
Finally, we address to the 2-dimensional Witten--Veneziano formula:
it yields a value for $F_{\eta}$, which we identify with  $F_{\pi}$,
as in large-$N_{\rm c}$ QCD. All three approaches consistently lead to
$F_{\pi} \simeq 1/\sqrt{2 \pi}$ at fermion mass $m=0$, which implies
that this quantity is meaningful.

\newpage

\tableofcontents

\section{Introduction}

The Schwinger model represents Quantum Electrodynamics in two
space-time dimensions \cite{Schwinger}.
This model shares several fundamental features
with 4-dimensional Quantum Chromodynamics (QCD), in particular
confinement \cite{Coleman} as well as the division of the gauge
configurations into topological sectors.

This model has been solved exactly in the massless case, but not
at finite fermion mass, $m>0$. In that case, analytic approaches
are usually based on bosonization and involve some assumptions
and approximations.

Here we consider the Schwinger model with $N_{\rm f} \geq 2$
degenerate fermion flavors, in Euclidean space-time.\\

Chiral Perturbation Theory is a systematic low-energy effective
theory of QCD, in terms of light meson fields.
Its Lagrangian includes a string of terms, which are
Lorentz invariant and chirally symmetric (if we refer to the chiral
limit, where the mesons are massless Nambu-Goldstone bosons).
The number of these terms is infinite, but they can be hierarchically
ordered in powers of the momenta, and truncated.
Each term has a coefficient, known
as a low-energy constant, which is a free parameter within Chiral
Perturbation Theory. It can only be determined from QCD as the
underlying, fundamental theory, or from experiment.

To leading order, there is only one term,
\be
{\cal L} = \frac{F_{\pi}^{2}}{4} \, \partial^{\mu} \vec \pi (x)
\cdot \partial_{\mu} \vec \pi (x) \ ,
\ee
where $\vec \pi$ is the pion field and the corresponding low-energy
constant $F_{\pi}$ is known as the {\em pion decay constant}.
It appears in a variety of relations, which are not necessarily
related to the pion decay.

Some of these relations occur in an analogous form in the multi-flavor
Schwinger model. Based on such analogies, we are going to discuss
three independent formulations of $F_{\pi}$ in the Schwinger model.
It is dimensionless in $d=2$, and the results obtained with these
three approaches are all compatible with the value
\be
F_{\pi} \simeq 1/\sqrt{2\pi} = 0.3989 \dots
\label{Fpivalue}
\ee
in the chiral limit.

In addition, this result is in good agreement with the only previous
determination that we are aware of: a study for $N_{\rm f}=2$ by
Harada {\it et al.} at strong coupling in a light-cone formulation
\cite{Harada}, which considered the relation
\be
\langle 0 | \partial^{\mu} J_{\mu}^{5}(0) | \pi (p)\rangle =
M_{\pi}^{2} F_{\pi} \ , 
\ee
where $J_{\mu}^{5}$ is the axial current and $M_{\pi}$ is
the ``pion'' mass. In this manner, Ref.\
\cite{Harada} obtained a mild dependence on the (degenerate)
fermion mass $m$,
\be  \label{FpiHarada}
F_{\pi}(m) = 0.394518(14) + 0.040(1) m/g \ , 
\ee
where $g$ is the gauge coupling, and $F_{\pi}(0)$ is close to our
value in eq.\ (\ref{Fpivalue}).

On the other hand, if one refers directly to the axial current,
instead of its divergence, 
\be
\langle 0 | J_{\mu}^{5}(0) | \pi (p)\rangle =
{\rm i} p_{\mu} F_{\pi} \ , 
\ee
one seems to arrive at $ F_{\pi} =0$, so the outcome does depend on
the QCD relation to which one establishes an analogy.

The QCD-inspired relations that we are going to refer to are the
Gell-Mann--Oakes--Renner relation (Section 2), the residual pion
mass in the $\delta$-regime (Section 3), and the Witten-Veneziano
formula (Section 4). Finally we present our conclusions
and an appendix about finite-size effects on $M_{\pi}$.
Preliminary results of this work were presented in a thesis
\cite{Jaime} and two proceeding contributions \cite{procs}.

\section{2d Gell-Mann--Oakes--Renner relation}

Back in 1992, Smilga derived the relation \cite{Smilga92}
\be
m \Sigma = C M_{\pi}^{2} \ ,
\ee
where $\Sigma$ is the chiral condensate, which --- in terms of
the fermion fields --- takes the usual form
$\Sigma = - \la \bar \Psi \Psi \ra$.
In the effective Lagrangian for QCD at small but non-zero quark
masses, $F_{\pi}$ and $\Sigma$ are the two leading low-energy
constants.

However, Smilga did not specify the constant $C$.
That was accomplished in Refs.\ \cite{HHI95,Hosotani,HosoRod}:
the bosonized 2-flavor Schwinger model leads to a Schr\"{o}dinger-type
equation, and in this framework these works studied the interactions of
(quasi) zero-modes due to the chiral anomaly and the fermion masses.
This led to an interesting formula (eq.\ (37) in Ref.\ \cite{HHI95}),
which --- in our notation and at zero vacuum angle --- reads
\be  \label{SigmaHHI}
\Sigma = \frac{M_{\pi}^{2}}{4\pi m} \ .
\ee
This relation is explained in detail in Ref.\ \cite{HosoRod}.
In addition, Ref.\ \cite{HHI95} also derived expressions for $M_{\pi}$
in terms of $m$, $g$ and the volume, in three different regimes. By
inserting $M_{\pi}$ into eq.\ (\ref{SigmaHHI}), the authors obtained
formulae for $\Sigma$ in each of these regimes.

However, that work did not relate eq.\ (\ref{SigmaHHI}) to the
``pion decay constant'', which we are interested in. This can be
achieved by invoking the Gell-Mann--Oakes--Renner relation \cite{GMOR},
which is well-known in QCD,
\be  \label{GMOR}
F_{\pi}^{2}(m) = \frac{2m}{M_{\pi}^{2}} \, \Sigma \ .
\ee
If we postulate the same relation in the multi-flavor Schwinger model,
and combine it with eq.\ (\ref{SigmaHHI}), we arrive at
\be
F_{\pi} = \frac{1}{\sqrt{2 \pi}} \ ,
\ee
without any mass dependence.\\

\begin{figure}[h!]
\vspace*{-1cm}
\begin{center}
  \includegraphics[angle=0,width=.8\linewidth]{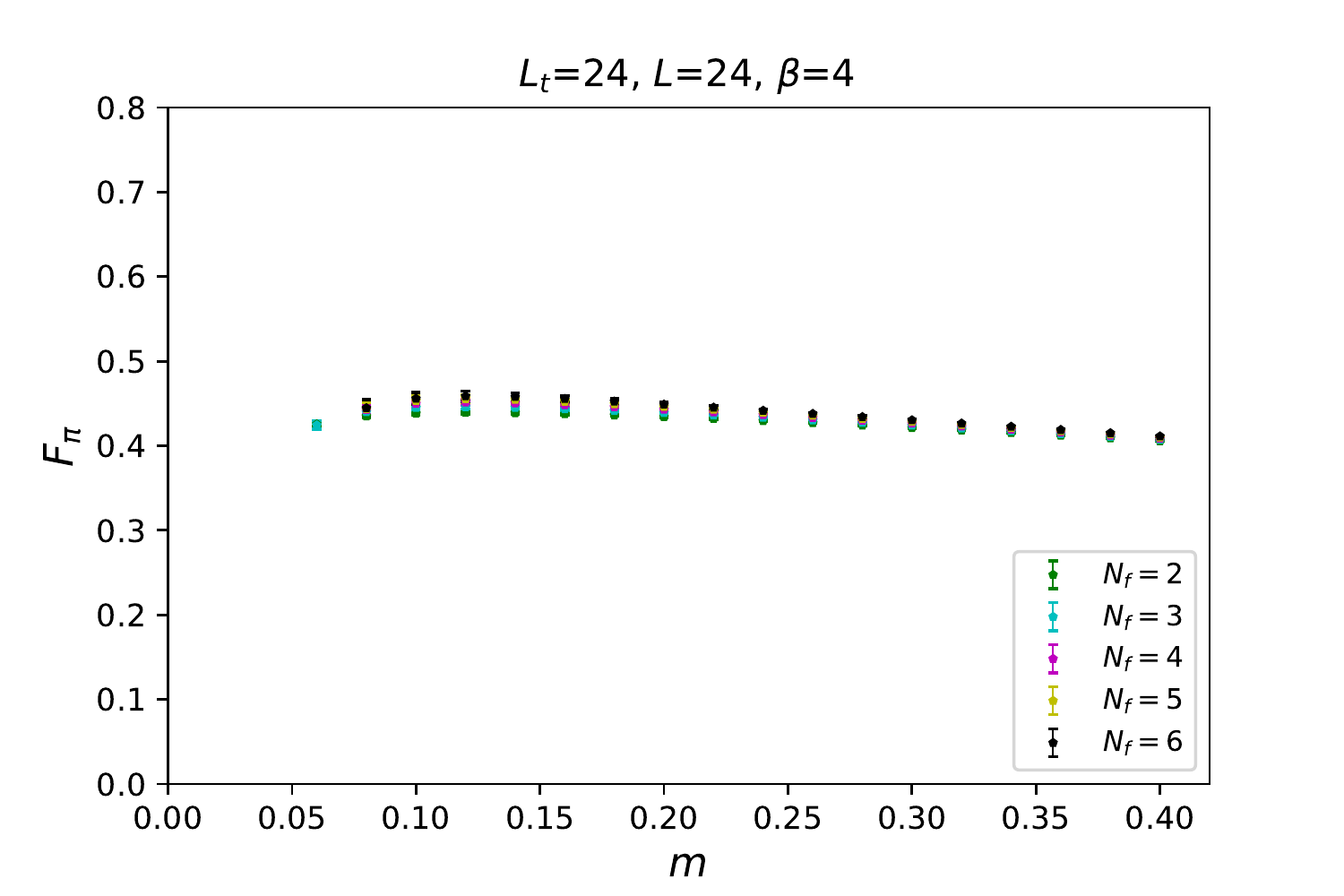}
  \includegraphics[angle=0,width=.8\linewidth]{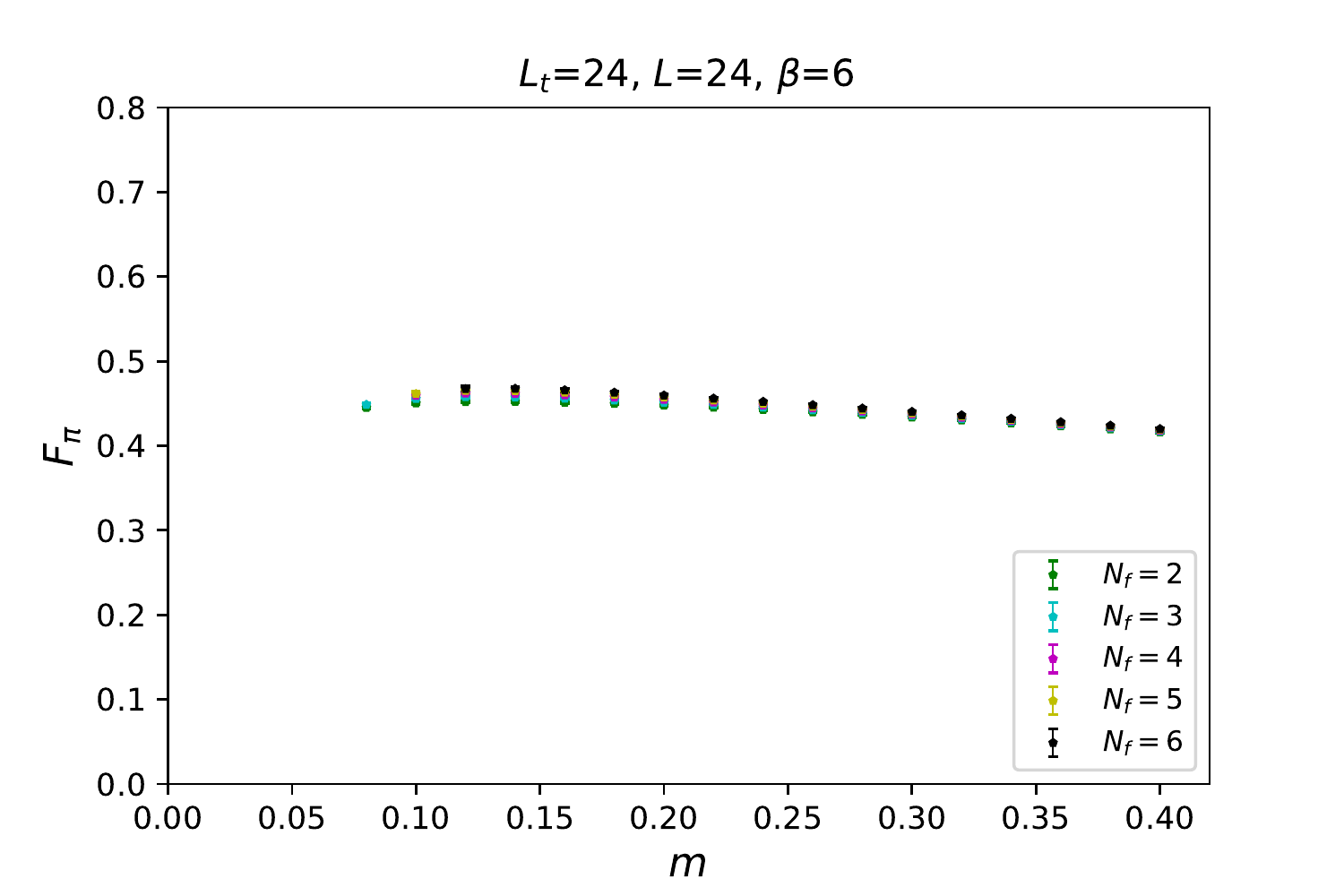}
\end{center}
\vspace*{-6mm}
\caption{\it Values for $F_{\pi}$ obtained from the Gell-Mann--Oakes--Renner
  relation (\ref{GMOR}) for $N_{\rm f}= 2, \dots , 6$ flavors,  
  at fermion masses $0.05 \leq m \leq 0.4$. The data are obtained from
  quenched simulations at $\beta=4$ (above) and at $\beta=6$ (below)
  with overlap-hypercube re-weighting, which works
  well, but for fermion mass $m \lesssim 0.05$ the results are
  affected by finite-size effects on $M_{\pi}$. We see convincing
  agreement for different $N_{\rm f}$, and hardly any difference
  for the different gauge couplings, hence the continuum limit
  seems smooth. In all cases, the extrapolations to the chiral limit
  are compatible with $F_{\pi} \simeq 0.4$.}
\label{FigGMOR1}
\end{figure}

Alternatively --- without relying on the approximations in the
bosonization approach --- we can numerically compute the quantities
on the right-hand side of eq.\ (\ref{GMOR}) in order to derive results
for $F_{\pi}(m)$. Such results are shown in Figure \ref{FigGMOR1}:
they were obtained based on quenched configurations on a lattice
of size $V= 24 \times 24$, generated at $\beta=4$ and $\beta=6$,
and re-weighting with the overlap-hypercube fermion determinant,
for the cases of $N_{\rm f} = 2, \dots , 6$ degenerate fermion flavors.
In Appendix A we are going to discuss the reliability of re-weighting
in such cases.

The overlap-hypercube Dirac operator is obtained by using the
overlap formula \cite{Neu}, which solves the Ginsparg-Wilson
relation \cite{Neu2}. This guarantees an exact, lattice-modified
chiral symmetry \cite{ML98}. However,
for the kernel we do not insert the usual Wilson operator, but
a truncated perfect hypercube fermion operator \cite{WB}. Compared
to the standard overlap formulation, this improves the scaling behavior,
approximate rotation invariance and the level of locality,
as demonstrated in quenched QCD \cite{WBQCD}.
The 2-dimensional version that we are using in the Schwinger model
was proposed in Ref.\ \cite{WBHip}, and applied also in
Refs.\ \cite{BHSV,LBH}.

Thanks to the chiral symmetry of the overlap-hypercube
operator, we can insert the bare fermion mass $m$, and reliably calculate
$M_{\pi}$ even at small $m$. $\Sigma$ is computed from the spectrum of
the Dirac operator,
\be  \label{Sigmam}
\Sigma (m) = \frac{1}{V} \left\la \sum_{k}
\frac{1}{\lambda_{k} + m} \right\ra \ ,
\ee
where the Dirac eigenvalues $\lambda_{k}$ are mapped from the
Ginsparg-Wilson circle (with center 1 and radius 1)
to the imaginary axis (their location
in the continuum limit) by means of a M\"{o}bius transform,
$\lambda_{k} \to \lambda_{k} /(1 - \lambda_{k}/2)$. 

Figure \ref{FigGMOR1} shows that the results for $F_{\pi}(m)$
are consistently in the magnitude of $0.4$.
With 30,000 (10,000) configurations at $\beta=6$ ($\beta=4$)
and for $m \gtapprox 0.2$
(in lattice units), the numerical values are quite precise.
This figure refers to the third regime in the case distinction of
eq.\ (36) in Ref.\ \cite{HHI95}, which is characterized (among other
conditions) by $L_{t} M_{\pi} \gg 1$ (in a volume $L \times L_{t}$).
We also observe consistent agreement in the range $\beta=4, \dots ,6$
and $N_{\rm f} = 2, \dots , 6$, which indicates that --- in this range
--- the value of $F_{\pi}(m)$ hardly depends on the gauge coupling and on
the number of flavors.

\begin{figure}[h!]
\vspace*{-2mm}
\begin{center}
\includegraphics[angle=0,width=.8\linewidth]{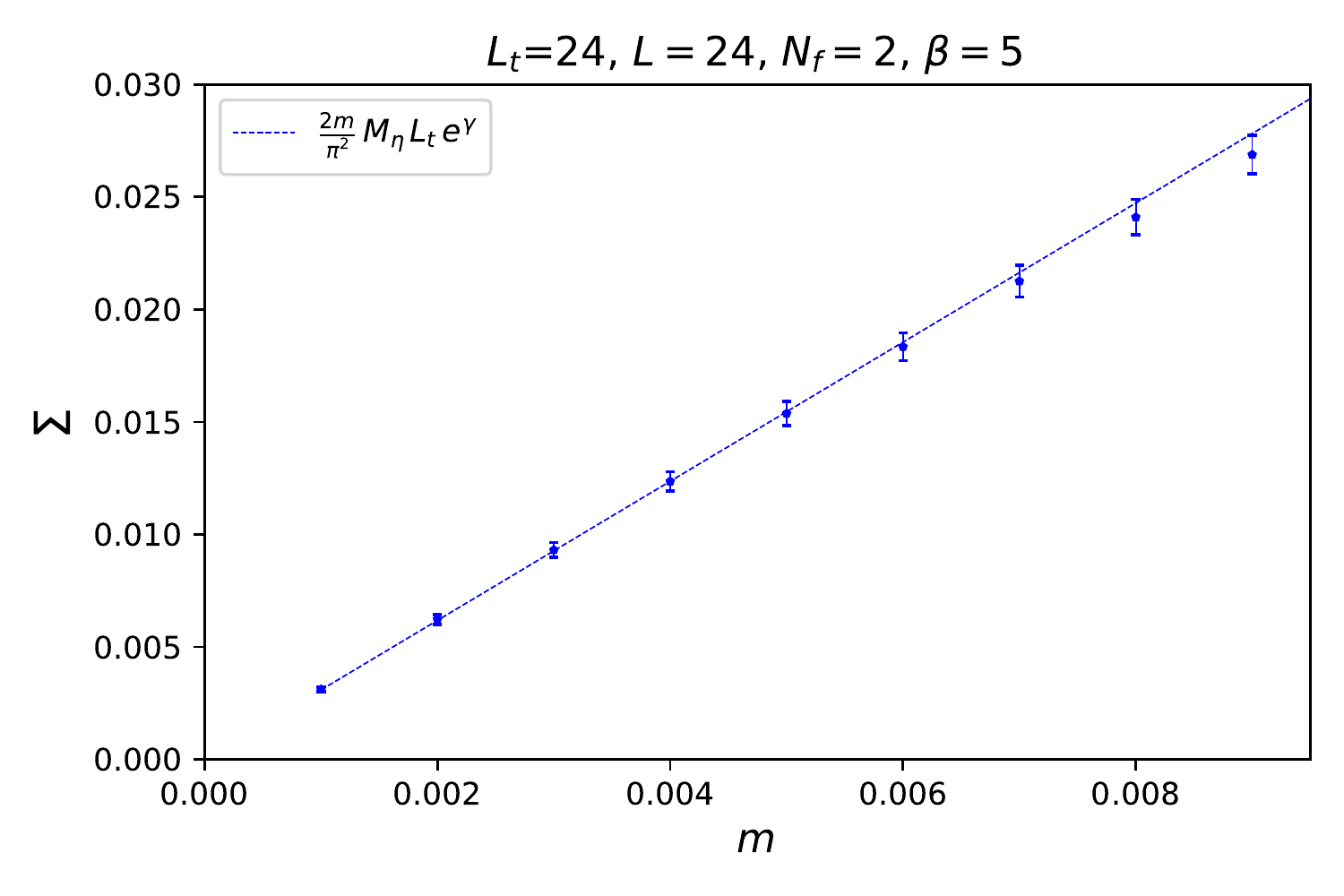}
\end{center}
\vspace*{-6mm}
\caption{\it The chiral condensate, measured for $N_{\rm f}=2$, at
$\beta=5$ on a $24 \times 24$ lattice, based on the Dirac spectrum
according to eq.\ (\ref{Sigmam}). It is compared to an asymptotic
formula for small $m$ given in Ref.\ \cite{HHI95}, where
$M_{\eta}$ represents the ``$\eta$-meson'' mass in the chiral
limit, see eq.\ (\ref{meta2}). We see that re-weighting works very
well even for fermion masses down to $m=0.001$ (cf.\  Appendix A), but
the Gell-Mann--Oakes--Renner relation for $F_{\pi}$, eq.\ (\ref{GMOR}),
also involves $M_{\pi}$, which is amplified by finite-size effects.}
\label{Sigmafig}
\vspace*{-2mm}
\end{figure}

At smaller fermion mass we enter the second regime of
eq.\ (36) in Ref.\ \cite{HHI95}, where $L_{t} M_{\pi} \ll 1$
(the spatial size $L$ remains large, so there is no relevant residual
pion mass due to finite size effects). Here the errors increase
visibly, and if $m$ is too small, even the
measured values are not reliable anymore: we still obtain good
results for $\Sigma$, as we see in Figure \ref{Sigmafig}, which
shows a comparison with predictions in Ref.\ \cite{HHI95}.
This also implies that re-weighting works well, at least for
$N_{\rm f} = 2$ flavors, even down to tiny fermion masses,
in agreement with earlier results in Ref.\
\cite{DurrHoelbling04}.\footnote{The reliability of re-weighting,
depending on $N_{\rm f}$ and $m$, will be further discussed in Appendix A.}
However, at tiny values of $m$, the pion mass $M_{\pi}$ suffers
from significant finite-size effects, since the product
$L M_{\pi}$ is not large anymore.
Moreover, in that regime there is a discrepancy between different
ways to measure $M_{\pi}$, as we will point out in Appendix B.

For all $N_{\rm f}$ that we included, we observe in Figure \ref{FigGMOR1}
a slight maximum of $F_{\pi}(m)$ between $m = 0.1$ and $0.15$.
At even smaller fermion mass $m$, $F_{\pi}(m)$ decreases and the
chiral extrapolation is again
compatible with $F_{\pi}(0) = 1/\sqrt{2 \pi}$, although at tiny $m$
the finite-size effects on $M_{\pi}$ and the large statistical errors
prevent a precise chiral extrapolation.

On the other hand, in some circumstances, the increase of $M_{\pi}$
due to finite-size effects can also be used to extract physical
information. This will be addressed in the next section.

\section{$F_{\pi}$ from the residual pion mass in the $\delta$-regime}

The approach of this section refers to Chiral Perturbation Theory,
which is a systematic effective field theory for low-energy QCD,
cf.\ Section 1.
One writes a general Lagrangian --- with all terms allowed by the
symmetries --- in terms of pseudo-Nambu-Goldstone boson fields.
In 2-flavor QCD, the fields represent the pions, which pick up
a small mass $M_{\pi}$ through non-zero masses of the $u$- and $d$-quark,
and by finite-size effects (if the volume is finite).

The latter are negligible in the most commonly used setting, the
$p$-regime of Chiral Perturbation Theory: here the space-time volume
is large, in all directions, compared to the correlation length $1/M_{\pi}$.
From a theoretical perspective, it is also instructive to study the
$\epsilon$-regime of a small space-time volume, and the $\delta$-regime,
with a small spatial box $L^3$ but a large extent $L_{t}$ in (Euclidean)
time, $L_{t} \gg L = {\cal O}(1/M_{\pi})$. In the $\epsilon$- and
$\delta$-regime, finite-size effects give rise to a significant
energy gap, hence the pions have a residual mass $M_{\pi}^{\rm R}$
even in the chiral limit of massless quarks.

Here we focus on the {\em $\delta$-regime:} it represents a
quasi-1-dimensional field theory, which formally corresponds to a
quantum mechanical system. Leut\-wyler introduced this regime
in Ref.\ \cite{Leutwyler}: he employed the picture of a quantum
mechanical rotor with the energy gap
\be  \label{mpiRgeneral}
M_{\pi}^{\rm R} = \frac{N_{\pi}}{2 \Theta} \ ,
\ee
where $N_{\pi}$ is the number of pions (or generally: of
pseudo-Nambu-Goldstone bosons) involved. The challenge is
to compute the ``moment of inertia'' $\Theta$. Leutwyler also
established the appropriate rules for the $\delta$-expansion,
and to leading order (LO) he obtained $\Theta = F_{\pi}^2 L^{3}$.

This expansion was extended to the next-to-leading order (NLO)
by Hasenfratz and Niedermayer, who referred to an O($N$) model
in $d>2$ dimensional Euclidean space \cite{Hasenfratz93}. According
to the Goldstone Theorem, the spontaneous symmetry breaking pattern
${\rm O}(N) \to {\rm O}(N-1)$ yields $N-1$ Nambu-Goldstone bosons,
which is the number to be inserted for $N_{\pi}$ in eq.\ (\ref{mpiRgeneral}),
along with
\be
\Theta = F_{\pi}^2 L^{d-1} \Big[ 1 + \frac{N_{\pi} -1}
{2 \pi F_{\pi}^2 L^{d-2}} \Big( \frac{d-1}{d-2}
- \frac{1}{2} \alpha_{1/2}^{(d-1)}(1)
\Big) \Big] \ ,
\ee
where $F_{\pi}$ has the mass dimension $d/2-1$.
(The constant $\alpha_{1/2}^{(d-1)}(1)$ is a shape coefficient;
its numerical values are given for symmetric boxes in various
dimensions in Ref.\ \cite{HasLeu}.)
Since that work refers to $d>2$, Nambu-Goldstone bosons are
present, and there was no problem with the pole at $d=2$.

Later the NNLO was investigated in Refs.\ \cite{Hasenfratz10,Niedermayer16}.
At this order, the sub-leading low-energy constants $l_{1}, \dots , l_{4}$
enter. Comparison of simulation data with the formulae of Ref.\ 
\cite{Hasenfratz10} yielded in particular
a sensible value for the controversial
coupling $l_{3}$ \cite{Bietenholz2010}. Another numerical study explored
the transitions from the $\delta$- to the $\epsilon$- and $p$-regime
\cite{Matzelle16}.

The current study refers to $d=2$, with a volume $L_{t} \times L$,
$L_{t} \gg L$. Here the Mermin-Wagner-Coleman
Theorem excludes Nambu-Goldstone bosons in the strict sense, but it
is known that the ``pions'' at small but finite fermion mass behave
similarly to pseudo-Nambu-Goldstone bosons in higher dimensions.
(At $m=0$ they decouple, thus avoiding a contradiction with the
Mermin-Wagner-Coleman Theorem \cite{SmiVer}.)
For considerations about the applicability of Chiral
Perturbation Theory in the Schwinger model, we refer to
Ref.\ \cite{KVZ}.

Due to the singularity at the NLO, we can only refer to the LO,
so we start from the hypothesis
\be  \label{mpiRd2}
M_{\pi}^{\rm R} = \frac{N_{\pi}}{2 F_{\pi}^{2} L} \ .
\ee
The basic prediction reduces to $M_{\pi}^{\rm R} \propto 1/L$,
which is plausible on dimensional grounds. If this is observed
numerically, we have another way to determine $F_{\pi}$, up to
the question how $N_{\pi}$ should be interpreted in this setting,
with $N_{\rm f}$ massless fermion flavors.

Part of the literature, for instance Refs.\ \cite{GatSei,Elser},
assumes $N_{\rm f}^2 -1$ ``pions''.
This matches the number of Nambu-Goldstone bosons in the spontaneous
symmetry breaking ${\rm SU}(N_{\rm f}) \otimes {\rm SU}(N_{\rm f}) \to
{\rm SU}(N_{\rm f})$ in $d \geq 3$ dimensions according to the
Goldstone Theorem. On the other hand, the literature which analyzes the
multi-flavor Schwinger model with bosonization usually deals with
$N_{\rm f} -1$ ``pions'' \cite{Belvedere,Affleck,HHI95,Hosotani,HosoRod}.

In fact, in the case $N_{\rm f}=2$ we obtain values for $F_{\pi}$,
which are consistent with the results based on the
Gell-Mann--Oakes--Renner relation, if and only if we insert $N_{\pi} = 1$.

When we proceed to $N_{\rm f}>2$, however, we see that the bosonization
formula $N_{\rm f} -1$ does not work anymore. So we take a pragmatic
point of view and adjust the number of ``pionic'' degrees
of freedom, which are manifest in formula (\ref{mpiRd2}). We obtain
consistent values for $F_{\pi}$, to an impressively high accuracy,
if we insert the effective formula
\be  \label{Npieffectiveformula}
N_{\pi} = \frac{2 (N_{\rm f} -1)}{N_{\rm f}} \ ,
\ee
although --- according to this formula --- $N_{\pi}$ is non-integer
for $N_{\rm f} \geq 3$.

\begin{figure}[h!]
\vspace*{-4mm}
\begin{center}
\includegraphics[angle=0,width=.8\linewidth]{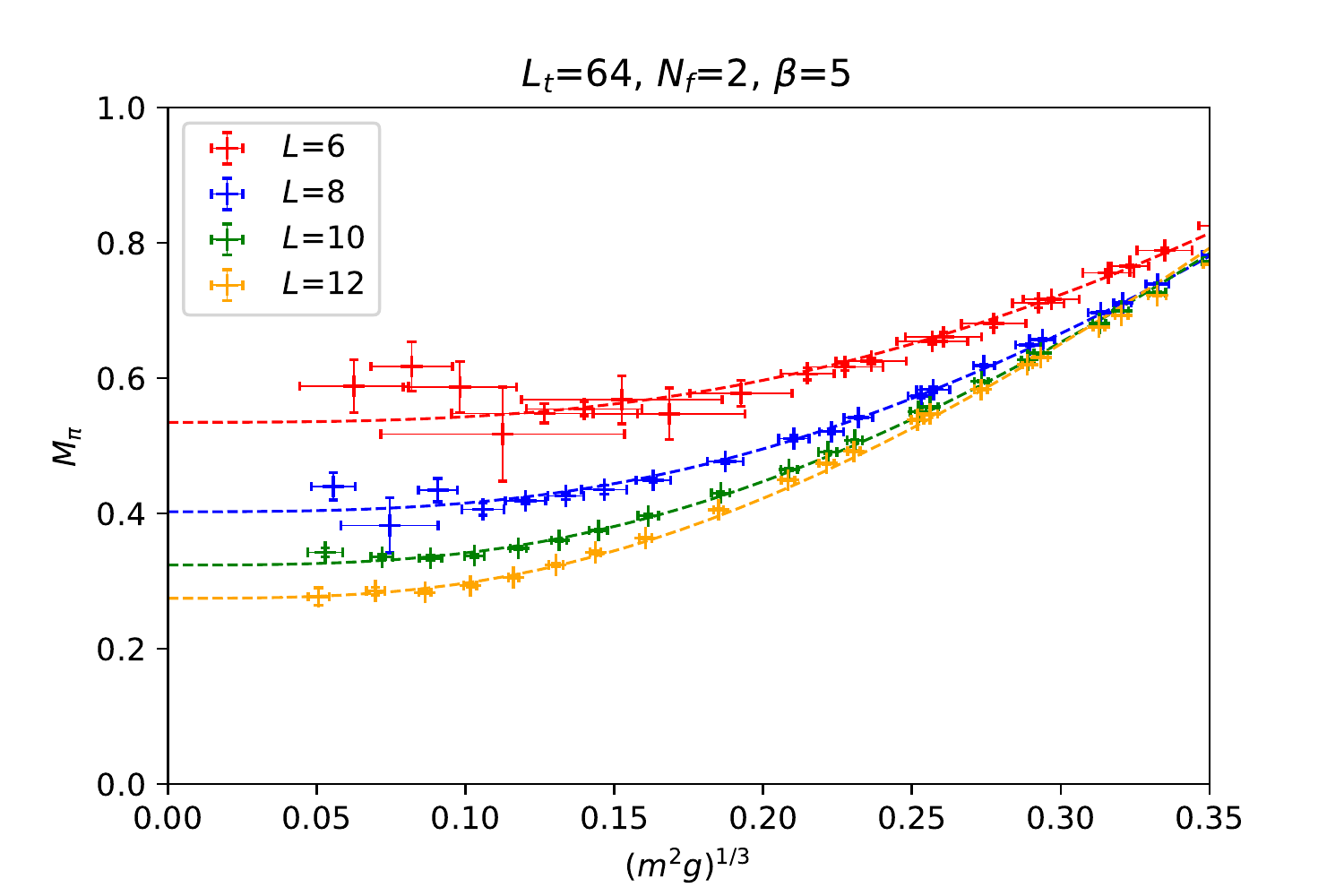}
\end{center}
\vspace*{-6mm}
\caption{\it Simulation results for the ``pion'' mass
  $M_{\pi}$ in the $\delta$-regime, $L \ll L_{t}$
  (with $L_{t} = 64$), using dynamical Wilson fermions.
  For a small fermion mass $m$ (determined by the PCAC relation)
  and a small spatial extent $L$, significant errors occur, as expected
  for Wilson fermions. Still, the full range of fermion masses enables
  sensible extrapolations to the residual ``pion'' mass $M_{\pi}^{\rm R}$
  in the chiral limit $m \to 0$.}
\label{DeltaRegWilsonMpivsm}
\end{figure}

\begin{figure}[h!]
\vspace*{-2mm}
\begin{center}
\includegraphics[angle=0,width=.8\linewidth]{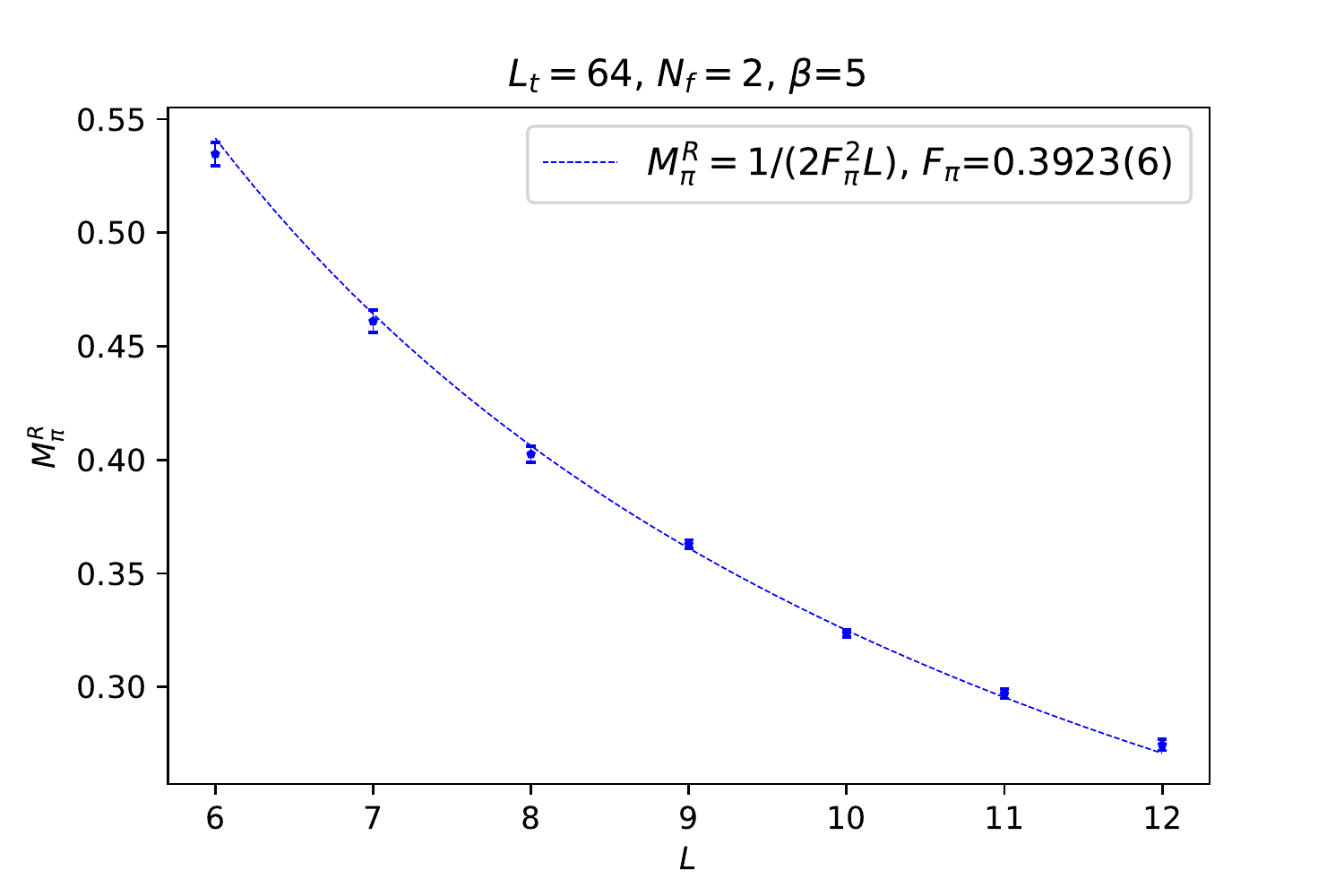}
\end{center}
\vspace*{-6mm}
\caption{\it The residual ``pion'' masses $M_{\pi}^{\rm R}$ in the
  $\delta$-regime, obtained from simulations
  of two flavors of dynamical Wilson fermions and extrapolated to the
  chiral limit according to Figure \ref{DeltaRegWilsonMpivsm}, in spatial
  volumes $L=6, \dots , 12$. The data follow well a fit proportional to $1/L$,
and the coefficient corresponds to $F_{\pi}=0.3923(6)$.}
\label{DeltaRegWilsonMpivsL}
\end{figure}

Let us substantiate this statement by presenting our simulation results.
We first refer to $N_{\rm f}=2$ flavors of dynamical Wilson fermions,
which are convenient to simulate. We set the Wilson parameter to 1
and used the Hybrid Monte Carlo algorithm \cite{HMC}, following the
scheme, which was established in Ref.\ \cite{GattHipLang} for
the 2-flavor Schwinger model: trajectories consist of 10 steps,
with the step-size being dynamically adjusted for a Metropolis
acceptance rate close to 0.8. We monitored the auto-correlations
of several observables --- including the topological charge ---
and separated the measurements by twice the maximal autocorrelation
time, in order to obtain practically de-correlated data sets. In this
way, we generated 10,000 configurations for each parameter set.

Of course, the results for Wilson fermions are plagued by
additive mass renormalization.
As usual for non-chiral lattice fermions, the renormalized fermion
mass $m$ is measured based on the PCAC relation.
Figure \ref{DeltaRegWilsonMpivsm} shows results for the ``pion''
mass $M_{\pi}$ in the $\delta$-regime, with $L_{t} = 64 \gg L$
($L=6,\, 8,\, 10,\, 12$),
which is plotted against the dimensionless parameter $(m^2 g)^{1/3}$, at
$\beta =1/g^{2} = 5$ (still in lattice units). As a generic property,
at decreasing, small values of $m$ and $(m^2 g)^{1/3}$, the statistical
errors of $M_{\pi}$ and in particular of $m$ itself increase
rapidly (at fixed statistics), but the complete set of results allows
for smooth fits with sensible extrapolations to the chiral limit $m=0$.

Figure \ref{DeltaRegWilsonMpivsL} shows these extrapolated values
of $M_{\pi}(m=0)$ as a function of the spatial size $L$ over the
range of $L=6, \dots , 12$. A fit confirms the expected behavior
$M_{\pi}(m=0) \propto 1/L$ to high accuracy, in particular up to
$L=11$. The proportionality constant is a fitting parameter,
which --- inserted in eq.\ (\ref{mpiRd2}) --- yields an
$F_{\pi}$-value close to the one in eq.\ (\ref{Fpivalue}),
$F_{\pi} = 0.3923(6)$.

We obtained very similar results at $\beta=3$ and $\beta=4$.
The corresponding figures are included in the first proceeding
contribution cited in Ref.\ \cite{procs}.
We do not reproduce these plots here, since they look almost
identical to Figures \ref{DeltaRegWilsonMpivsm} and
\ref{DeltaRegWilsonMpivsL}, but we display the $F_\pi$-values
obtained in this manner at three gauge couplings in Table
\ref{tabFpidelta} (the value are slightly modified due
to improved data analysis). They coincide to percent level
thus providing clear evidence that the continuum limit is again
very smooth, as we observed before in the consideration of
Section 2.
\begin{table}[h!]
 \begin{center}
  \begin{tabular}{|c||c|c|c|}
        \hline   
	$\beta = 1/g^{2} $ & 3 & 4 & 5 \\
        \hline
        $F_\pi$  & 0.3887(7) & 0.3877(11) & 0.3923(6) \\
	\hline
  \end{tabular}
 \end{center}
 \vspace*{-2mm}
 \caption{Results for $F_{\pi}$, obtained by fits to eqs.\
(\ref{mpiRd2}) and (\ref{Npieffectiveformula}), with $N_{\rm f}=2$,
at three values of $\beta$.}
 \label{tabFpidelta}
 \vspace*{-2mm}
\end{table}

Next we proceed to results that we obtained with overlap-hypercube fermions,
by using 10,000 gauge configurations that we generated
quenched\footnote{The quenched configurations used here and in
  other sections were generated with the same HMC algorithm,
  by setting the fermion determinant to a constant.}
at $\beta=4$,
which were re-weighted again for the case of $N_{\rm f}=2$ flavors.
\begin{figure}[h!]
\vspace*{-2mm}
\begin{center}
\includegraphics[angle=0,width=.8\linewidth]{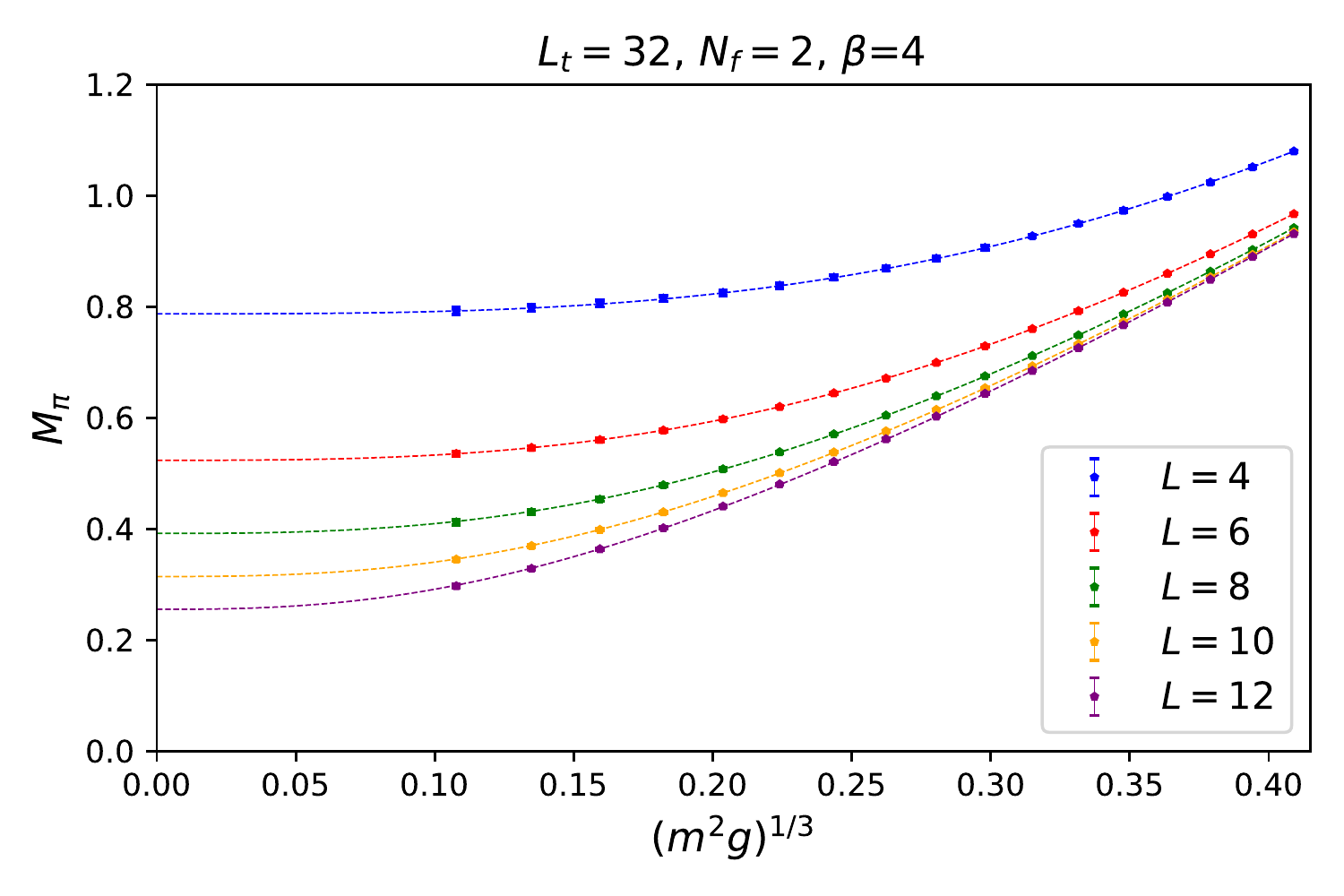}
\end{center}  
\vspace*{-6mm}
\caption{\it Like Figure \ref{DeltaRegWilsonMpivsm}, but here the ``pion''
  mass $M_{\pi}$ is measured with overlap-hypercube fermions, using
  quenched, re-weighted gauge configurations, generated at $\beta=4$.
  In contrast to Figure \ref{DeltaRegWilsonMpivsm}, this yields small
  errors and smooth chiral extrapolations for all spatial sizes
  $L=4, \dots , 12$ under consideration.}
\label{DeltaRegOverlapMpivsm}
\end{figure}
As we see in Figure \ref{DeltaRegOverlapMpivsm}, the exact, lattice
modified chirality of the overlap-hypercube fermions strongly suppresses
the statistical fluctuations at relatively
small fermion mass $m$, which --- in this case --- is directly taken from
the Lagrangian. Thus in this approach the values for $M_{\pi}(m=0)$ are
quite precise. They represent the residual ``pion'' mass in the
$\delta$-regime with $L_{t} =32$ and $L = 4, \dots , 12$.
For a further discussion of the re-weighting approach, we refer again
to Appendix A.

\begin{figure}[h!]
\vspace*{-2mm}
\begin{center}
\includegraphics[angle=0,width=.8\linewidth]{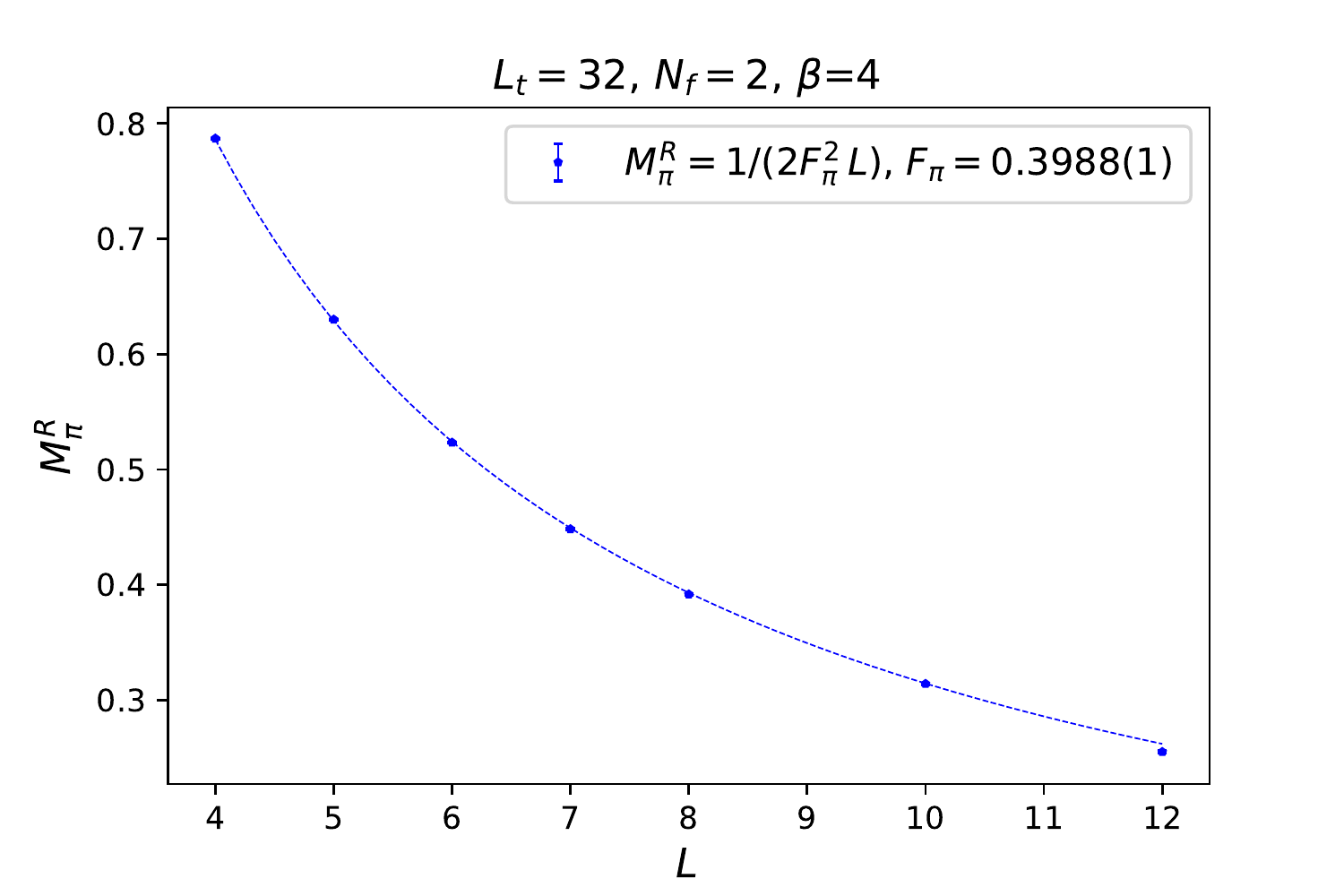}
\end{center}
\vspace*{-6mm}
\caption{\it Like Figure \ref{DeltaRegWilsonMpivsL}, but now with data
  obtained from the extrapolation of overlap-hypercube fermions results,
  see Figure \ref{DeltaRegOverlapMpivsm}. Again the fit to the conjectured
  behavior $M_{\pi}^{\rm R} \propto 1/L$ works very well for $L<12$, and
  we extract $F_{\pi}=0.3988(1)$. This value is well compatible with further
  results that we obtained for $F_{\pi}$ by employing different methods,
  and in perfect agreement with formula (\ref{Fpivalue}).}
\label{DeltaRegOverlapMpivsL}
\end{figure}

Figure \ref{DeltaRegOverlapMpivsL} shows that these safely extrapolated 
values again follow very well a behavior $M_{\pi}(m=0) \propto 1/L$,
at least for $L<12$. Here the fitting constant leads to
\be
F_{\pi} = 0.3988(1) \ ,
\ee
in remarkable proximity to the result obtained with Wilson fermions,
and in perfect agreement with formula (\ref{Fpivalue}).

\begin{figure}[h!]
\vspace*{-2mm}
\begin{center}
\includegraphics[angle=0,width=.8\linewidth]{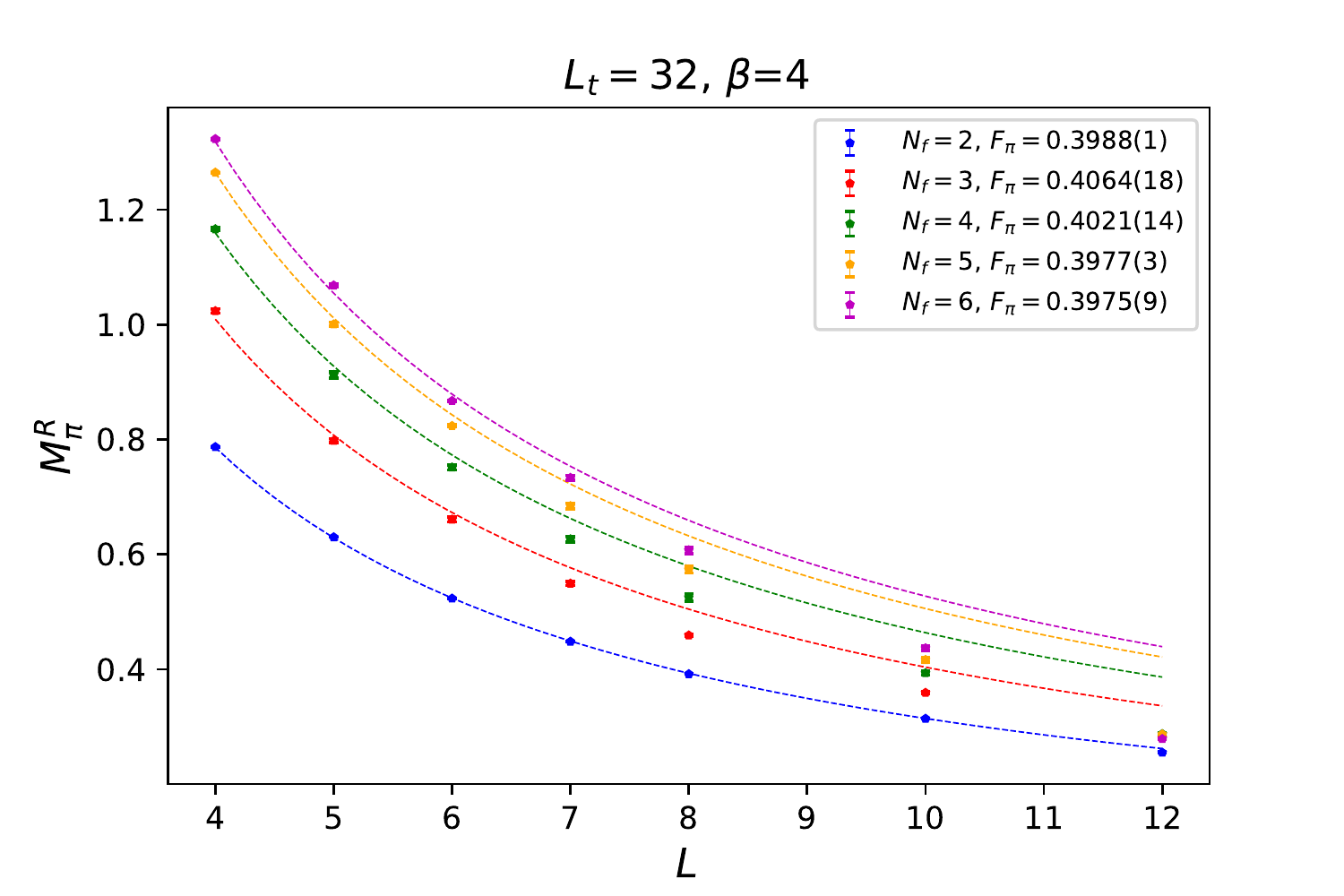}
\end{center}
\vspace*{-6mm}
\caption{\it Residual ``pion'' masses $M_{\pi}^{\rm R}$ in the $\delta$-regime
  ($L_t = 32$) for a variety of spatial sizes $L \ll L_{t}$, and
  $N_{\rm f}= 2, \dots , 6$ flavors. We show chiral extrapolations of
  quenched, re-weighted results with overlap-hypercube fermions,
  at $\beta =4$.
  The fits were performed in the range where they are successful,
  {\it i.e.}\ in the full range for $N_{\rm f}=2$, and for $L \leq 6$
  for $N_{\rm f}>2$. They lead to highly consistent values for $F_{\pi}$,
  if we apply the effective formula (\ref{Npieffectiveformula}).}
\label{DeltaRegOverlapMpivsLNf2to6}
\end{figure}

Finally, we extend the study with quenched and overlap-hypercube
re-weighted configurations up to $N_{\rm f} = 6$ degenerate flavors,
as in Section 2. Figure \ref{DeltaRegOverlapMpivsLNf2to6} illustrates the
residual ``pion'' masses against the spatial lattice size $L=4, \dots , 12$.
As $N_{\rm f}$ increases, the behavior $\propto 1/L$ is observed
only up to $L=6$; at somewhat larger $L$, the residual
``pion'' mass stays below this proportionality relation.

However, when we restrict the fit to the range where the relation
$M_{\pi}^{R} \propto 1/L$ is well approximated,
we consistently obtain $F_{\pi} = 0.399(1)$ over
this range of $N_{\rm f}$, if we insert the effective formula
(\ref{Npieffectiveformula}). This underscores that the value
$F_{\pi} = 1/\sqrt{2 \pi}$ is meaningful, and that eq.\
(\ref{Npieffectiveformula}) correctly captures the number of
``pionic'' degrees of freedom which are manifest in the
$\delta$-regime (even if this number is non-integer).
We add that any attempts to extend the fits $\propto 1/L$ at
large $N_{\rm f}$ up to larger spatial size $L$ lead to an unsatisfactory
fitting quality (a modified power of $L$ would be required),
hence they do not provide alternative $F_{\pi}$-results.

\section{Witten-Veneziano formula in the Schwinger model}

The Witten-Veneziano formula is well-known in the framework of
QCD \cite{WitVen}: it refers to the 't Hooft large-$N_{\rm c}$ limit,
which keeps the product $g_{\rm s}^{2} N_{\rm c}$ constant ($g_{\rm s}$ is
the strong gauge coupling and $N_{\rm c}$ the number of colors). This
limit overcomes the axial anomaly of chiral 3-flavor QCD. Hence the
spontaneous symmetry breaking pattern takes the form
${\rm U}(3)_{\rm L} \otimes {\rm U}(3)_{\rm R} \to {\rm U}(3)_{\rm L=R}$
(where the subscripts L and R denote the quark chiralities),
and we obtain a nonet of Nambu-Goldstone bosons: they correspond to
the pions, the kaons and the mesons $\eta$ and $\eta '$,
which are all massless in this limit.

The Witten-Veneziano formula expresses the mass that the $\eta '$-meson
picks up due to the leading $1/N_{\rm c}$-corrections. For the more
general case of $N_{\rm f}$ massless quark flavors, this mass is
given by
\be
M_{\eta '}^{2} = \frac{2 N_{\rm f} \chi_{\rm t}^{\rm q}}{F_{\eta '}^{2}} \ ,
\ee
where $\chi_{\rm t}^{\rm q}$ is the quenched topological susceptibility,
which can be measured by means of lattice simulations. In this
particular case, the quenched value is relevant,
because quark loops do not contribute to this order in the 
$1/N_{\rm c}$-expansion. Moreover, in this order the pion decay
constant coincides with the $\eta'$-decay constant,
\be
F_{\pi} = F_{\eta '} \ .
\ee
Inserting the experimental value of $F_{\pi} \simeq 92.4~{\rm MeV}$
and simulation results for $\chi_{\rm t}^{\rm q}$, see in particular
Ref.\ \cite{WuppChit}, (roughly) confirms the
observed mass $M_{\eta '} \simeq 958 \, {\rm MeV}$. Thus the fact that
$\eta'$ is far heavier than the light meson octet (and even a little
heavier than a nucleon) is explained as a topological effect.
This is the quantitative solution to the U(1) problem.

\begin{figure}[h!]
\begin{center}    
  \includegraphics[angle=0,width=.8\linewidth]{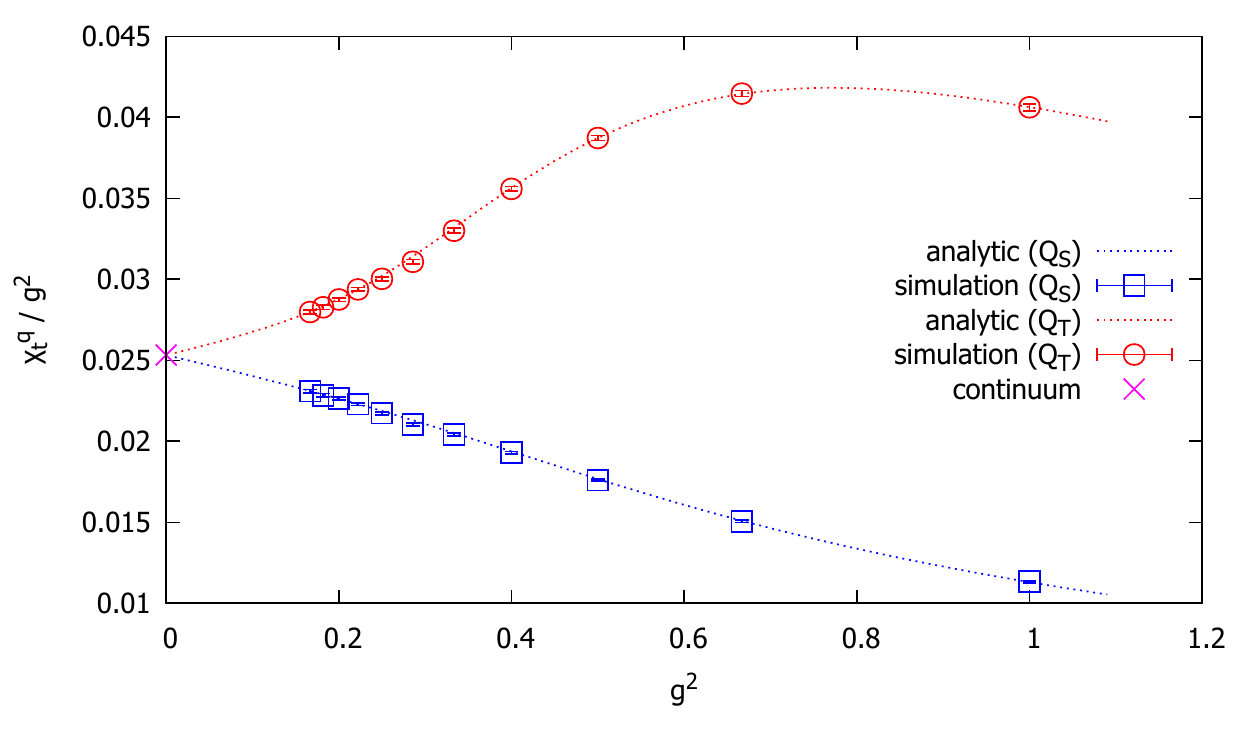}
\end{center}
\vspace*{-6mm}
\caption{\it The quenched topological susceptibility $\chi_{\rm t}^{\rm q}$,
  for two different lattice formulations of the topological charge density
  (standard plaquette term $\theta_{\rm P}$ and $\sin (\theta_{\rm P})$).
  In the latter case, the topological charge
  $Q_{\rm S} = \sum_{P} \sin (\theta_{\rm P}) /2\pi$ can be computed
  analytically \cite{BDEH}, while in case of
  $Q_{\rm T} = \sum_{P} \theta_{\rm P} /2\pi$ the sum over the
  plaquettes can be computed numerically \cite{BonatiRossi}.
  In both cases, the values are in excellent agreement with
  our simulation results.
  Both formulations consistently lead to the continuum limit with
  $\chi_{\rm t}^{\rm q} / g^{2} = 1/4 \pi^{2}$, which was derived in
  Ref.\ \cite{SeiSta}.}
\label{topsusquenched}
\end{figure}

According to Seiler and Stamatescu, the Witten-Veneziano relation is
actually on more solid grounds in the framework of the Schwinger model
with $N_{\rm f} \geq 2$ massless fermion flavors \cite{SeiSta}.
In the chiral limit, it takes the form
\be  \label{SchwingVW}
M_{\eta}^{2} = \frac{2 N_{\rm f}}{F_{\eta}^{2}} \chi_{\rm t}^{\rm q} \ ,
\ee
where the ``$\eta$-meson'' is the meson-type singlet state.
Its mass has been computed analytically \cite{Belvedere},
\be  \label{meta2}
M_{\eta}^{2} = \frac{1}{\pi} N_{\rm f} g^{2} \ .
\ee
Ref.\ \cite{SeiSta} further derived the following relation
for the quenched, topological susceptibility (in the continuum and
infinite volume)
\be  \label{SeilerStamtopsus}
\chi_{\rm t}^{\rm q} = \frac{g^{2}}{4 \pi^{2}} \ .
\ee
Figure \ref{topsusquenched} shows results for
$\chi_{\rm t}^{\rm q}/g^{2}$ obtained for two
lattice formulations of the topological charge,
\be
Q_{\rm T} = \sum_{P} \theta_{\rm P} /2\pi \ , \quad 
Q_{\rm S} = \sum_{P} \sin (\theta_{\rm P}) /2\pi \ .
\ee
The sums run over all plaquettes $P$, and
$\theta_{\rm P}$ is the plaquette discretization of the topological
density $\epsilon_{\mu \nu} \partial_{\mu} A_{\nu}$. $Q_{\rm T} \in \Z$
is the standard formulation, which can be numerically evaluated
to high precision (see {\it e.g.} Ref.\ \cite{BonatiRossi}).
For the alternative formulation $Q_{\rm S}$,
the lattice topological charges are in general non-integer, but
Ref.\ \cite{BDEH} derived an analytic formula at finite $g$,
{\it i.e.}\ at finite lattice spacing, in terms of Bessel functions,
$\beta \chi_{\rm t}^{\rm q} = I_{1}(\beta) / [4\pi^{2} I_{0}(\beta)]$.
In both cases, we computed $\chi_{\rm t}^{\rm q}$
at finite $g$ also with Monte Carlo simulations. The results
agree accurately, and the continuum limit smoothly leads to the
value given in eq.\ (\ref{SeilerStamtopsus}), for both formulations,
as Figure \ref{topsusquenched} shows. This result is also in agreement
with Ref.\ \cite{DurrHoelbling04}.

Inserting eqs.\ (\ref{meta2}) and (\ref{SeilerStamtopsus}) into
eq.\ (\ref{SchwingVW}), we obtain
\be
F_{\eta} = \frac{1}{\sqrt{2 \pi}} \ .
\ee
At this point, we push the analogy to large-$N_{\rm c}$ QCD further
and assume $F_{\pi} = F_{\eta}$.
We are not aware of a basic justification of this step,
but it exactly confirms once more formula (\ref{Fpivalue}).

\section{Summary and conclusions}

In this work, we have attracted attention to a dimensionless
constant, which plays a relevant role in the multi-flavor Schwinger
model, but which has been ignored in most of the literature. The
only exception is a study by Harada {\it et al.}\ \cite{Harada} in
the light-cone formulation, which led to the result that we quoted
in eq.\ (\ref{FpiHarada}).

By analogy to specific aspects of QCD, we denote this constant as
$F_{\pi}$, as it was done before in Ref.\ \cite{Harada}. We derived
its value by another three independent methods, which all provide
consistent results. In particular, referring to the 2d
Gell-Mann--Oakes-Renner relation and inserting formulae from
bosonization approaches \cite{Smilga92,HHI95,Hosotani,HosoRod} leads
to $F_{\pi} = 1/\sqrt{2 \pi}$, which is also compatible with
simulation results. The residual ``pion'' mass in the
$\delta$-regime confirms this value to a good precision, if we
rely on relations of Chiral Perturbation Theory even in the absence
of Nambu-Goldstone bosons, and on our effective formula
(\ref{Npieffectiveformula}) for the
number of light degrees of freedom. Finally, the Witten-Veneziano
formula yields $F_{\eta} = 1/\sqrt{2 \pi}$, and if we identify
$F_{\pi} = F_{\eta}$, as in large-$N_{\rm c}$ QCD, we arrive once
more at the same value for $F_{\pi}$.

The first method seems most robust. The latter two involve some
{\it ad hoc} assumptions, which are, however, motivated from
analogies to QCD. The impressive agreement of the results for
$F_{\pi}$ cannot be by accident, so we conclude that these
{\it ad hoc} assumptions are --- in this context --- sensible.
This concerns in particular our effective formula
(\ref{Npieffectiveformula}) for the ``pionic'' degrees of freedom,
which are manifest in the $\delta$-regime, as well as the relation
$F_{\pi} = F_{\eta}$. It further implies
that the constant $F_{\pi} = 1/\sqrt{2 \pi}$ is indeed relevant
for the multi-flavor Schwinger model, in particular for the case
$N_{\rm f}=2$. The underlying reason, as well as further
appearances of $F_{\pi}$ in the Schwinger model, remain to be
explored. \\

\noindent
{\bf Acknowledgments:} We thank Stephan D\"{u}rr, Christian Hoelbling
and Satoshi Iso for helpful comments.
The code was developed at the cluster Isabella of the Zagreb University
Computing Centre (SRCE), and the production runs were performed at
the cluster of the Instituto de Ciencias Nucleares, UNAM.
This work was supported by the Faculty of Geotechnical Engineering
of Zagreb University through the project ``Change of the Eigenvalue
Distribution at the Temperature Transition'' (2186-73-13-19-11),
by UNAM-DGAPA through PAPIIT project IG100322,
``Materia fuertemente acoplada en condiciones extremas con el MPD-NICA'',
and by the
Consejo Nacional de Humanidades, Ciencia y Tecnolog\'{\i}a (CONAHCYT).

\appendix

\section{Reliability of re-weighting with an overlap
  fermion determinant}

Our results presented in Section 2, and part of the results in Section 3,
were obtained with $10^{4}$ or $3 \times 10^{4}$
quenched configurations for each setting, which were re-weighted
with the fermion determinant of the overlap-hypercube Dirac operator.
In this appendix, we discuss the reliability of this procedure by
decomposing the contributions to the chiral condensate $\Sigma$
(as an example), according to formula (\ref{Sigmam}).
To this end, the contributions are summed up in hierarchical order.
The question is how many of the configurations, which are
dominant in this respect, are needed to arrive at a good
approximation of our value for $\Sigma$ based on the
entire statistics, {\it i.e.}\ how many of these configurations
are statistically relevant.

\begin{figure}[h!]
\begin{center}
\includegraphics[angle=0,width=.527\linewidth]{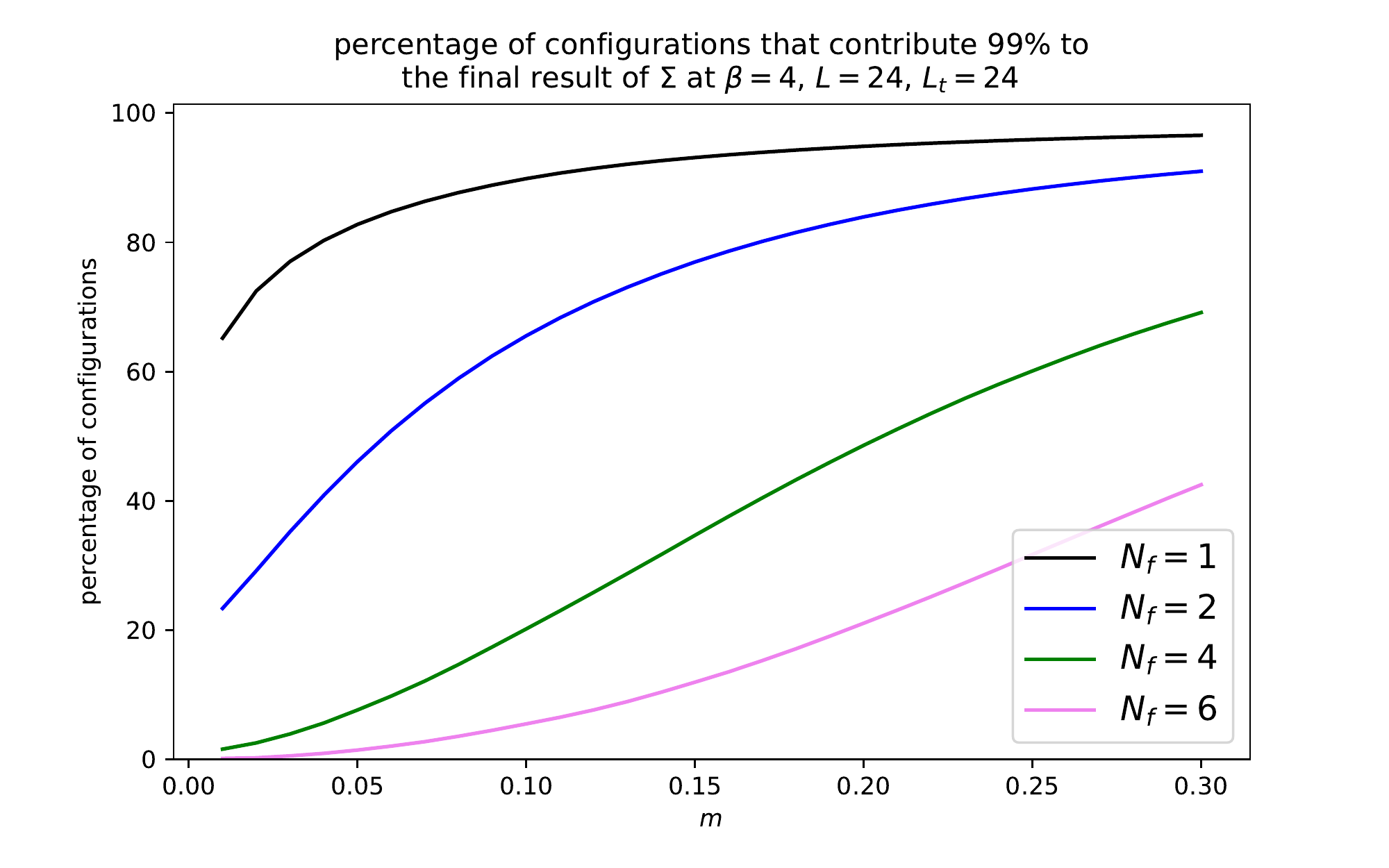}
\hspace*{-10mm}
\includegraphics[angle=0,width=.527\linewidth]{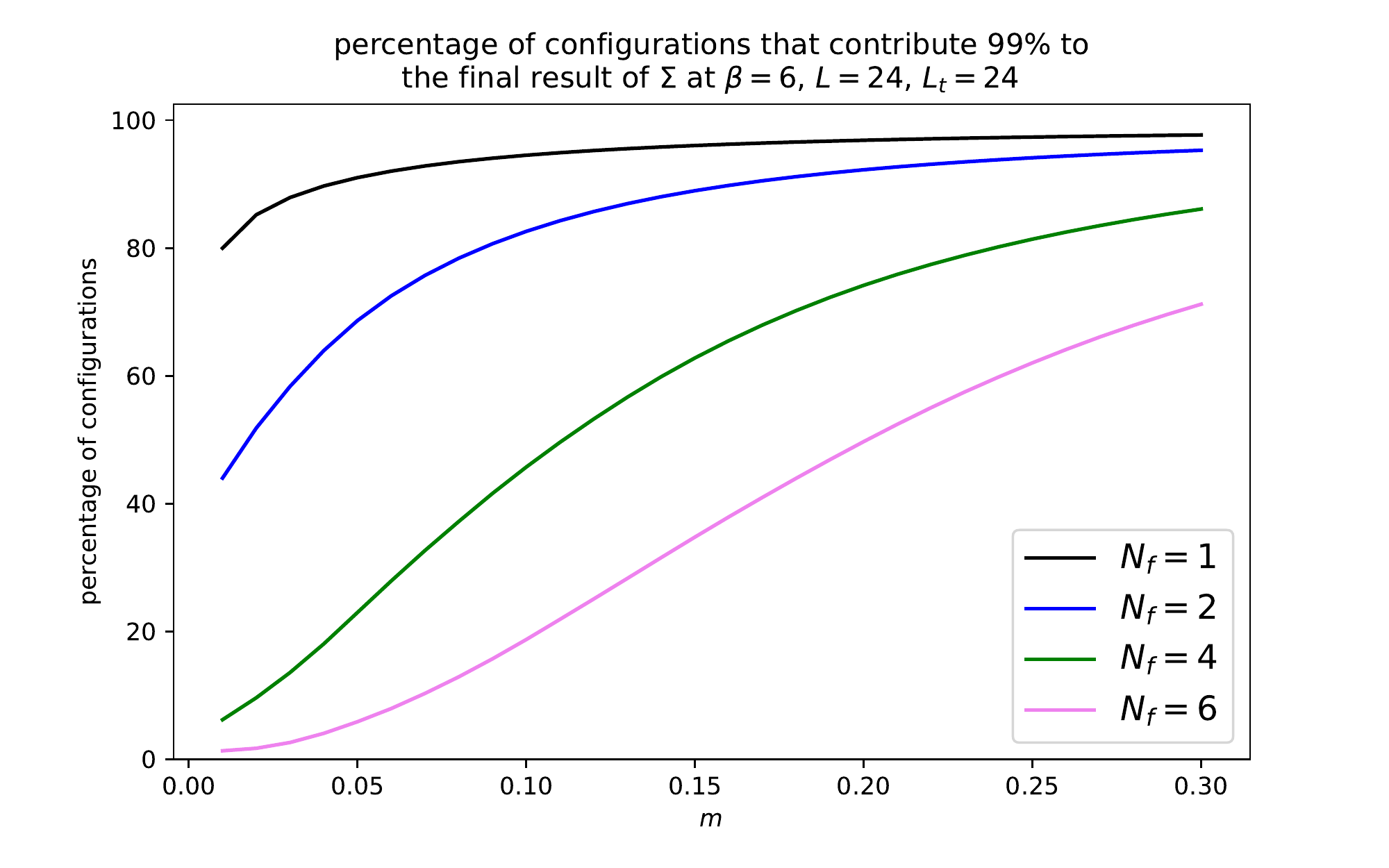}
\end{center}
\vspace*{-6mm}
\caption{\it The number of leading contributions which have to be included
  to obtain $99\,\%$ of our value for the chiral condensate $\Sigma$,
  at $\beta=4$ (left) and $\beta =6$ (right), on a $24 \times 24$
  lattice.
  For small fermion mass $m$ and a relatively large number $N_{\rm f}$
  of flavors, only a minor part of the configurations is necessary
  and therefore relevant.}
\label{ContriSigma}
\end{figure}

In Figure \ref{ContriSigma} we consider the percentage of dominant
configurations, which is sufficient to obtain our total value
of $\Sigma$ up to $1\,\%$.
We illustrate both the dependence on the (degenerate) fermion
mass $m$ and on the number of flavors $N_{\rm f}$, {\it i.e.}\
the power of the fermion determinant.
For low $N_{\rm f}$ and moderate values of $m$, a large fraction
of the contributions is needed to obtain $99\,\%$ of our
$\Sigma$-value, hence the effective statistics is not much below
the total data set. Vice versa, for increasing $N_{\rm f}$ and small
$m$, the results are essentially just due to a minor subset of
configurations, hence the {\em effective statistics} is substantially
reduced compared to the total statistics.
Still, we saw in Figure \ref{Sigmafig} that at $N_{\rm f}=2$
and $\beta=5$ re-weighting works well even down to $m=0.001$,
where the effective statistics is about
$30\,\%$ of the data set. On the other hand, a large number of
flavors can drastically suppress the effective statistics.
For instance, at $m \lesssim 0.1$ we see that $N_{\rm f}>6$ would be
worrisome indeed, hence we do not show any result for even more
flavors than $N_{\rm f} = 6$ in this work.

\begin{figure}[h!]
\begin{center}
\includegraphics[angle=0,width=.527\linewidth]{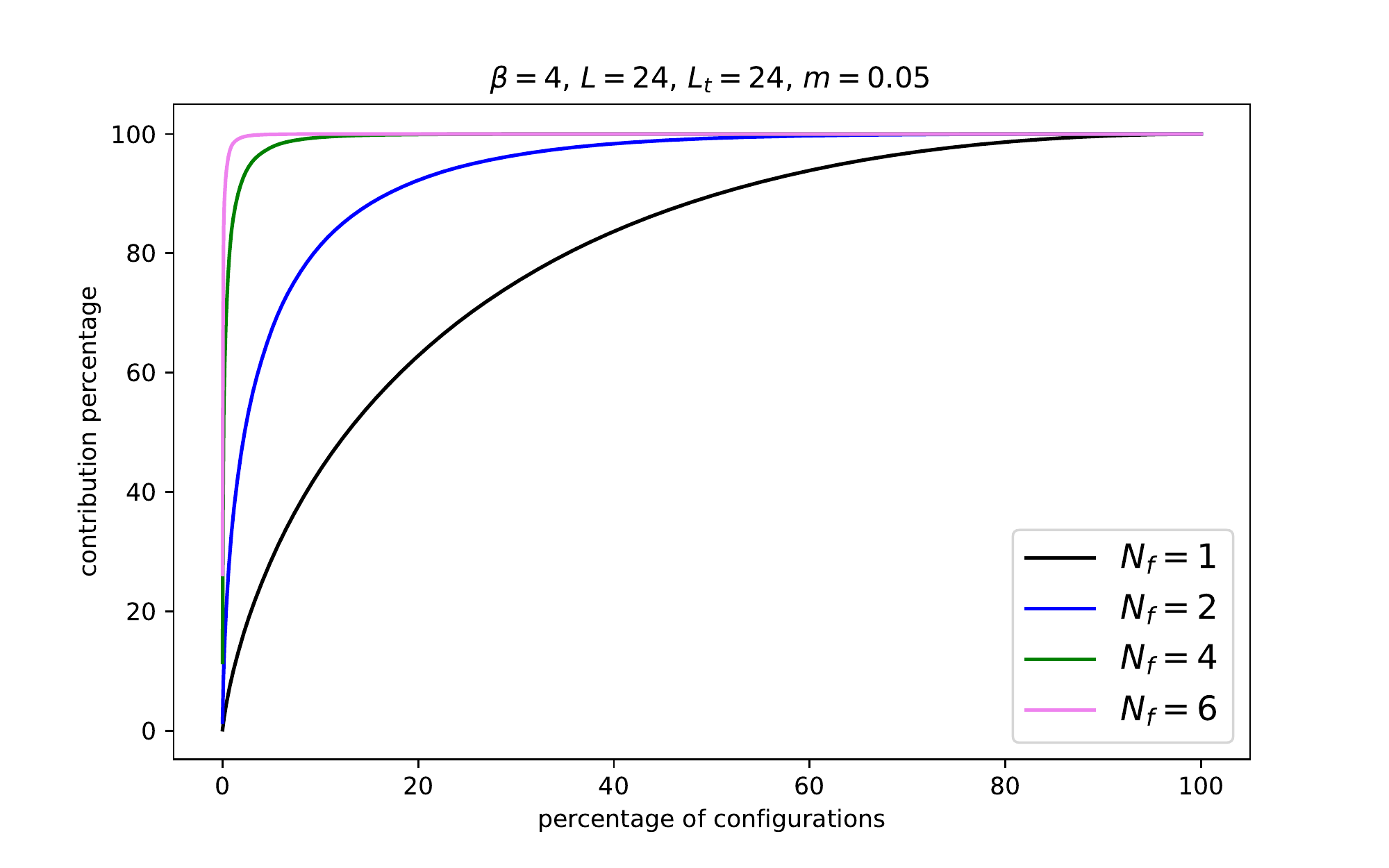}
\hspace*{-10mm}
\includegraphics[angle=0,width=.527\linewidth]{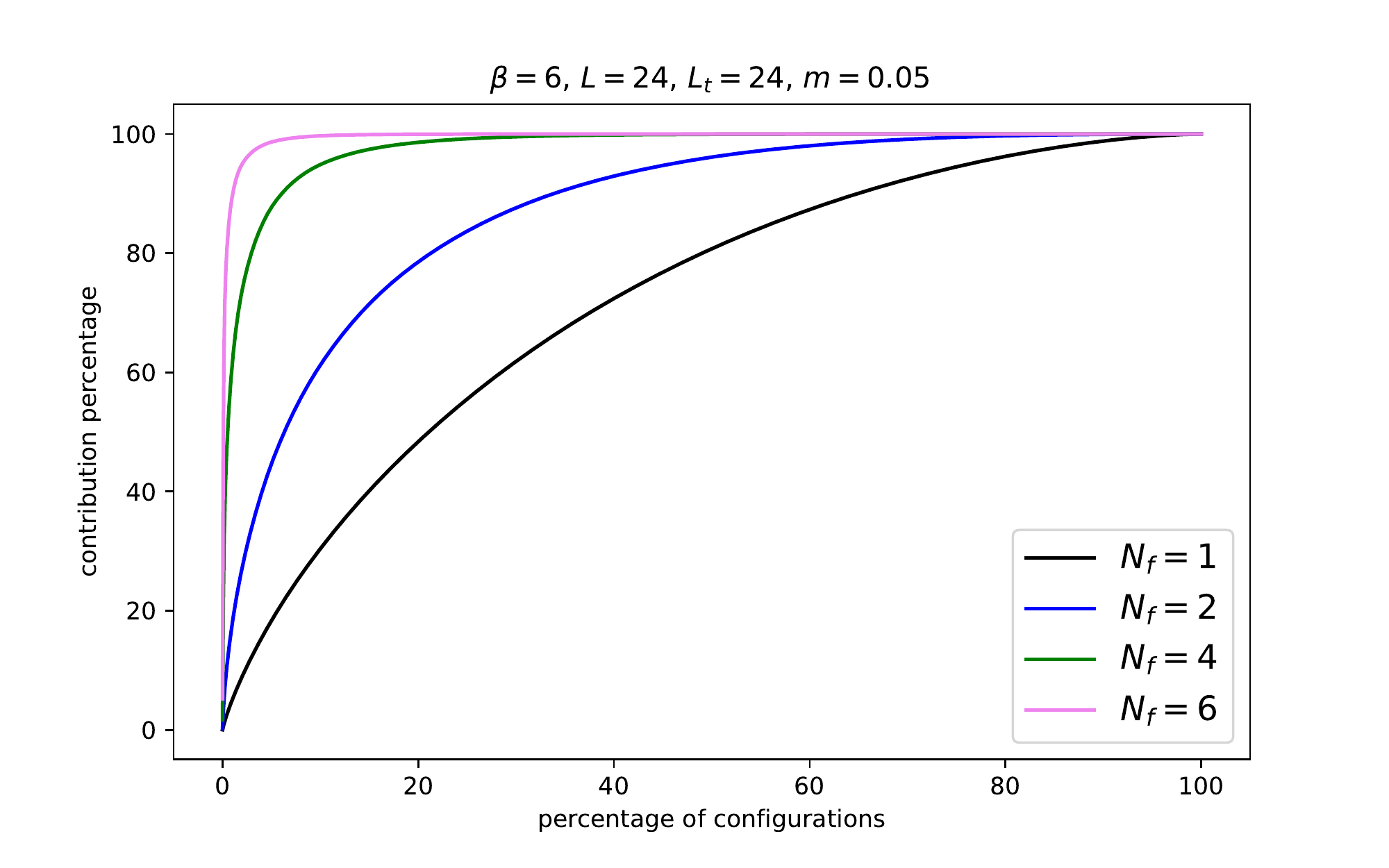}
\end{center}
\vspace*{-6mm}
\caption{\it The percentage of our $\Sigma$-value as a function of the
  number of dominant contributions, out of a total of $10^{4}$ configurations
  at $\beta=4$ (left) and $3 \times 10^{4}$ configurations at $\beta =6$
  (right), at $m=0.05$ on a $24 \times 24$ lattice. We see that even
  more flavors than $6$ would be troublesome.}
\label{CumulativeSigma}
\end{figure}

Part of our results in Sections 2 and 3 included a minimal
fermion mass of $m=0.05$. For that particular mass, Figure
\ref{CumulativeSigma} shows the percentage of our $\Sigma$-value
as a function of the number of leading contributions. Again we see
that for a considerable $N_{\rm f}$, in particular for $N_{\rm f} = 6$,
most of our result is due to only few contributions, which
confirms the limitation of the re-weighting method.

\section{The ``pion'' mass in the $\epsilon$-regime}

In Sections 2 and 4 we showed simulation results obtained
on $L\times L$ square lattices. In these cases, the measured ``pion''
mass $M_{\pi}$ is close to its value in the thermodynamic limit
($L \to \infty$), since the condition $L \gg 1/M_{\pi}$ is reasonably
well approximated.
\begin{figure}[h!]
\vspace*{-2cm}
\begin{center}
\includegraphics[angle=0,width=.8\linewidth]{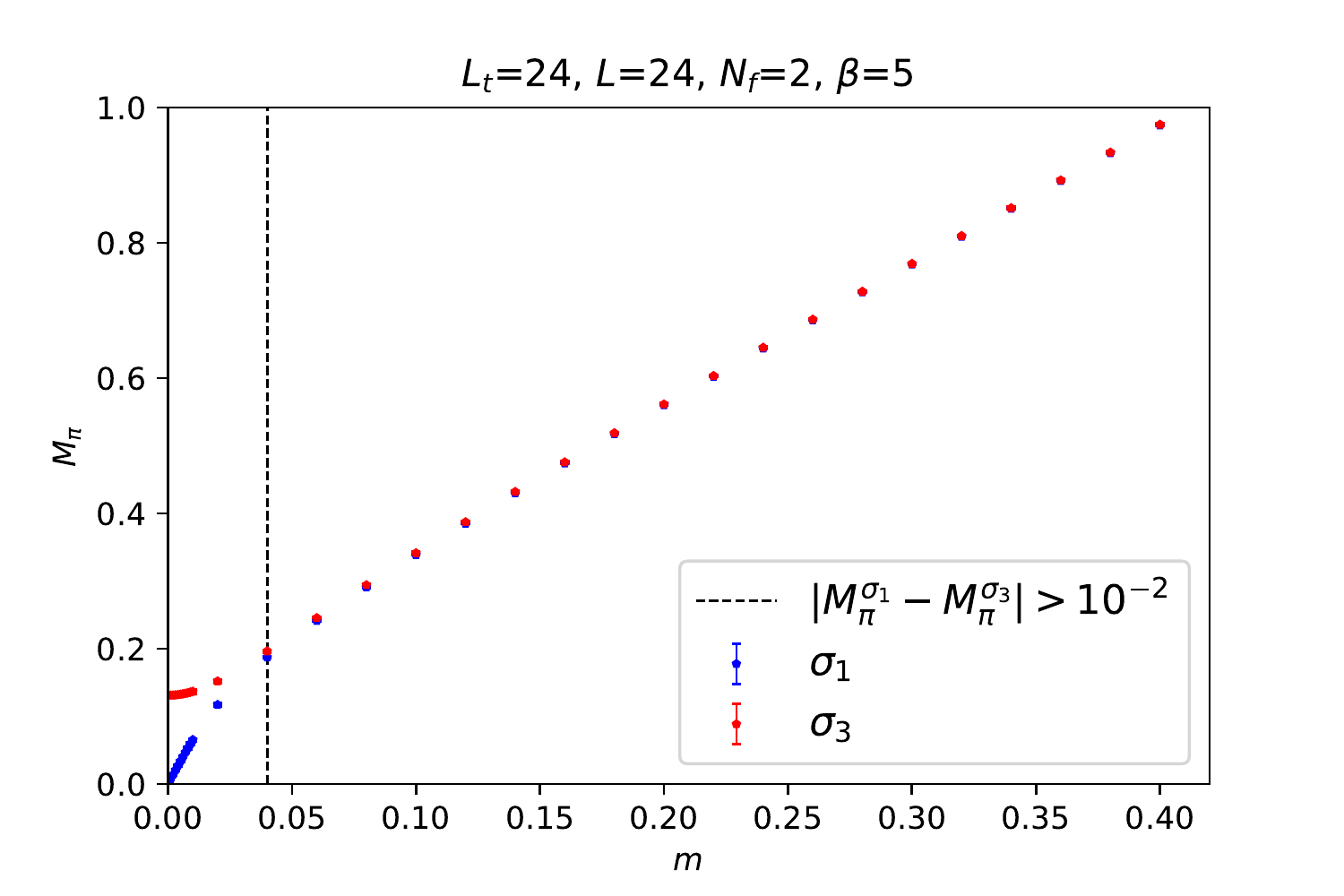}
\includegraphics[angle=0,width=.8\linewidth]{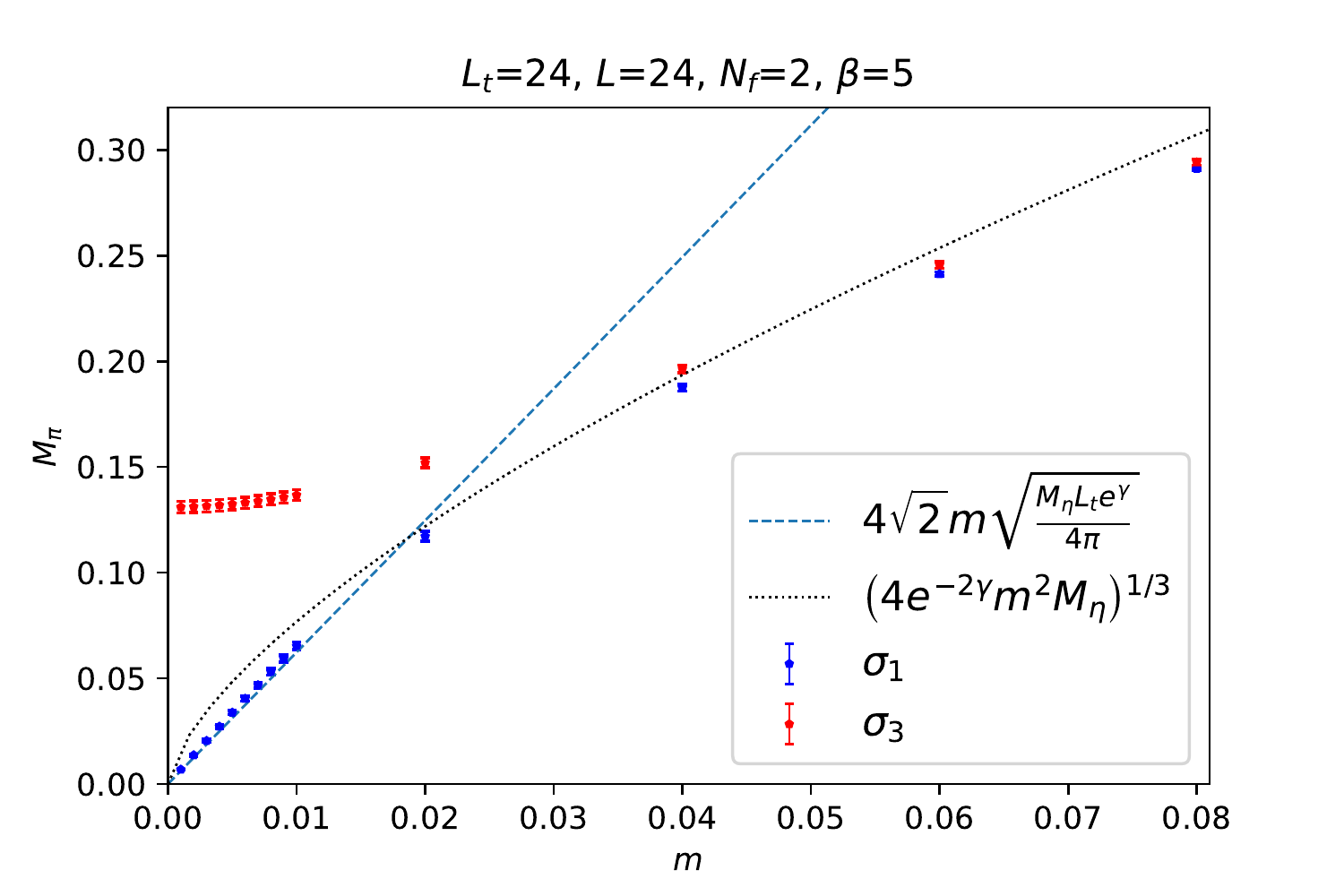}
\end{center}
\vspace*{-6mm}
\caption{\it Illustration of the ``pion'' mass measured with $\sigma_{1}$
($M_{\pi}^{\sigma_{1}}$) and with $\sigma_{3}$ ($M_{\pi}^{\sigma_{3}}$),
for $N_{\rm f}=2$, at $\beta=5$ on a $24 \times 24$ lattice.
The upper plot shows that there is good agreement at fermion mass
$m \geq 0.1$, but at $m \leq 0.05$ they differ drastically:
$M_{\pi}^{\sigma_{3}}$ attains a residual value, which is
the expected $\epsilon$-regime behavior,
whereas $M_{\pi}^{\sigma_{1}}$ drops to $0$ with an approximately
linear dependence on $m \ltapprox 0.02$. The lower plot zooms into the
small-$m$ region, and compares the data to eq.\ (36) of Ref.\
\cite{HHI95}, where $M_{\eta} = g \sqrt{2/\pi}$ is the ``$\eta$-mass''
in the chiral limit, cf.\ eq.\ (\ref{meta2}), and $\gamma$ is Euler's
constant.
For $m \gtapprox 0.05$, $M_{\pi}^{\sigma_{1}}$ and $M_{\pi}^{\sigma_{3}}$ are
close to each other, and to the prediction in the third regime of Ref.\
\cite{HHI95}.}
\label{BreakingPoint}
\end{figure}
Down to the corresponding values for the fermion
mass $m$, we also observed agreement of $M_{\pi}$ calculated either
with the correlation function of the density $\bar \psi \sigma_{3} \psi$,
or with $\bar \psi \sigma_{1} \psi$. As usual, we refer to a Dirac operator
in terms of $\sigma_{1}$ and $\sigma_{2}$, and both formulations
have been used in the literature. The former is closer to the concept
of the physical pion, but the latter is a valid alternative
in the range of the plots in Sections 2 and 4.

However, the situation changes when we proceed to even smaller values
of $m$. Here we enter the $\epsilon$-regime, where it is natural that
$M_{\pi}$ is significantly enhanced by finite-size effect.
Moreover, we observed that these two formulations of $M_{\pi}$ react
very differently to the squeezing in a small physical volume, as we
illustrate in Figure \ref{BreakingPoint}.

For the formulation with $\sigma_{3}$, one obtains a plateau with
a residual ``pion'' mass $M_{\pi}^{\sigma_{3}}$, similarly to the
$\delta$-regime, which is the generic behavior.
For the $\sigma_{1}$-formulation, however,
$M_{\pi}^{\sigma_{1}}$ approaches 0, closely following the relation
$M_{\pi}^{\sigma_{1}} \propto m$, which is an artifact due to the use
of $\sigma_{1}$.
In this sense the $\sigma_{1}$-formulation is a valid alternative
only in large volumes.

However, it is an amazing observation that $M_{\pi}^{\sigma_{1}}$ at tiny
$m \ltapprox 0.02$ accurately follows the prediction in eq.\ (36) of
Ref.\ \cite{HHI95} in the second regime, where $M_{\pi}L_{t} \ll 1$.
We referred to it before in Section 2; in that prediction, there is
no residual ``pion mass'' because Ref.\ \cite{HHI95} deals with a
large spatial size $L$. This is not the setting of our simulation,
but $M_{\pi}^{\sigma_{1}}$ follows this prediction to high accuracy.
The reason for this observation remains to be explored.

As the fermion mass increases to $m \gtapprox 0.05$, the ``pion
masses'' measured in both ways are close, $M_{\pi}^{\sigma_{1}} \approx
M_{\pi}^{\sigma_{3}}$, and they are now in the vicinity of the third
regime eq.\ (36) of Ref.\ \cite{HHI95}, where $M_{\pi} L \gg 1$,
and $M_{\pi} \propto m^{2/3}$.

This is a technical observation, which could be of interest
for future lattice studies, but it does not seem to be
documented in the literature.

\end{document}